\documentclass[epj]{svjour}
\usepackage{graphics}
\usepackage{amsmath}
\usepackage[normalem]{ulem} 
\usepackage{color}

\newcommand{\be}{\begin{eqnarray}}
\newcommand{\ee}{\end{eqnarray}}
\newcommand{\ba}{\begin{array}}
\newcommand{\ea}{\end{array}}

\newcommand{\nablaboth}{\stackrel{\leftrightarrow}{\nabla}}
\renewcommand\sout{\bgroup \color{blue} \ULdepth=-.5ex \ULset}

\begin{document}
\title{3D Structure and Nuclear Targets}
\author{Rapha\"el Dupr\'e \inst{1} \and Sergio Scopetta \inst{2}} 
\institute{Institut de Physique Nucl\'eaire, 
CNRS/IN2P3 and Universit\'e Paris Sud, Orsay, France \and 
Dipartimento di Fisica e Geologia, Universit\`a degli Studi
di Perugia, via A. Pascoli 06100 Perugia, Italy and INFN, sezione di Perugia
}
\date{Received: date / Revised version: date}
%
\abstract{
Recent experimental and theoretical ideas are laying the ground for a new era 
in the knowledge of the parton structure of nuclei. We report on two promising 
directions beyond inclusive deep inelastic scattering experiments, aimed at, 
among other goals, unveiling the three dimensional structure of the bound 
nucleon. The 3D structure in coordinate space can be accessed through deep 
exclusive processes, whose non-perturbative content is parametrized in terms 
of generalized parton distributions. In this way the distribution of partons 
in the transverse plane will be obtained, providing a pictorial view of the 
realization of the European Muon Collaboration effect. In particular, we 
show how, through the generalized parton distribution framework, non nucleonic 
degrees of freedom in nuclei can be unveiled. Analogously, the momentum space 
3D structure can be accessed by studying transverse momentum dependent parton 
distributions in semi-inclusive deep inelastic scattering processes. The 
status of measurements is also summarized, in particular novel coincidence 
measurements at high luminosity facilities, such as Jefferson Laboratory. 
Finally the prospects for the next years at future facilities, such as the 
12~GeV Jefferson Laboratory and the Electron Ion Collider, are presented.  
} 
\maketitle
\section*{Introduction}
\label{intro}

The nucleus is a unique laboratory for fundamental studies of the QCD hadron 
structure. For example, the extraction of the neutron information from light 
nuclei, essential for a precise flavor separation of parton distributions 
(PDs), the measurement of nuclear PDs, relevant for the analysis of 
nucleus-nucleus scattering aimed at producing quark-gluon plasma, or the 
phenomenon of in-medium fragmentation, mandatory to unveil the dynamics of 
hadronization, require nuclear targets. Nevertheless, inclusive Deep Inelastic 
Scattering (DIS) off nuclei has proven to be unable to answer a few 
fundamental questions. Among them, we list: (i) the quantitative microscopic 
explanation of the so called European Muon Collaboration (EMC) 
effect~\cite{Aubert:1983xm}, i.e., the medium 
modification of the nucleon parton structure; (ii) the full understanding of 
the structure of the neutron; (iii) the medium modification of the 
distribution of parton transverse momentum, relevant for studies of 
hadronization as well as of chiral-odd quantities, such as the transversity 
PDs or the Sivers and Collins functions.

Novel coincidence measurements at high luminosity facilities, such as 
Jefferson Laboratory (JLab), have become recently possible, 
addressing a new era in the 
knowledge of the parton structure of nuclei~\cite{dupre_scope}.
In particular, two promising directions beyond 
inclusive measurements, aimed at unveiling the three dimensional (3D) 
structure of the bound nucleon, are deep exclusive processes off nuclei,
and semi-inclusive deep inelastic 
scattering (SIDIS) involving nuclear targets,
In deep exclusive processes, one accesses the 3D structure in coordinate space, in terms
of generalized parton distributions (GPDs)~\cite{gpds_here};
in SIDIS, the momentum space 3D 
structure can be obtained by studying transverse momentum dependent parton 
distributions (TMDs)~\cite{gpds_here}.
In the following, we show how, in this way, a relevant contribution is expected to the 
solution of long standing problems, such as: (i) the non nucleonic contribution
to nuclear structure, (ii) the quantitative explanation of the 
medium modification of the nucleon parton structure, (iii) a precise flavor 
separation of the nucleon parton distributions, 
or (iv) the mechanism of in-medium 
hadronization as a fundamental test of confinement.

The report is structured as follows. The next section is dedicated to show one 
of the first motivations for the measurement of nuclear GPDs, i.e., how the 
contribution of non-nucleonic degrees of freedom can be singled out, while the 
same contributions are much more difficult to be accessed in standard DIS 
experiments~\cite{Berger:2001zb}. In the second section, another idea in favor 
of the measurements of nuclear GPDs, proposed in%
~\cite{Polyakov:2002yz}, will be reported, together with its most recent 
developments. Thanks to this proposal, using an interesting relation between 
GPDs and one of the
form factors of the parton energy momentum tensor, the 
spatial distribution of shear-forces experienced by the partons in the nucleus 
could be experimentally accessed. In the third section, the general issue of 
modifications of nucleon GPDs in the nuclear environment will be reported. In 
the fourth section, the possibility to use light nuclear targets to have a 
flavor separation of GPDs and TMDs is described. The fifth section is 
dedicated to the modification of parton transverse momentum in nuclei, to be 
studied through SIDIS and the TMD framework, in particular 
to its interplay 
with the nuclear transport parameter measured in hadronization experiments. 
Conclusions are eventually drawn in the final section.

\section{Non-nucleonic degrees of freedom in nuclei from nuclear GPDs}
\label{sec:1}

The first paper on nuclear GPDs~\cite{Berger:2001zb}, concerning the 
deu\-te\-ron, contained already the crucial observation that the knowledge of 
GPDs would permit the investigation of the interplay of nucleon and parton 
degrees of freedom in the nuclear wave function. In standard DIS off a nucleus 
with four-momentum $P_A$ and $A$ nucleons of mass $M$, this information can be 
accessed in the region where $x_{Bj} = \frac{Q^2}{2 M \nu}$ is greater 
than 1, $\nu$ being the energy transfer in the laboratory system and $Q^2$ the 
momentum transfer. In this region, kinematically forbidden for a free proton 
target, very fast quarks are tested and measurements are therefore very 
difficult, because of vanishing cross-sections. As explained 
in~\cite{Berger:2001zb,Cano:2003ju}, a similar information can be accessed in 
DVCS at lower values of $x_{Bj}$.

To understand this aspect, it is instructive to analyze coherent DVCS in 
Impulse Approximation (IA). Let us think, to fix the ideas, to unpolarized 
DVCS off a nucleus with $A$ nucleons, which is sensitive to the GPD $H_q^A$ 
only. This has been treated in \cite{Cano:2003ju} for the deuteron target, 
in~\cite{Guzey:2003jh} for spin-0 nuclei, in~\cite{Kirchner:2003wt} for nuclei 
with spin up to 1, in~\cite{Scopetta:2004kj} for $^3$He and 
in~\cite{Liuti:2005gi} for $^4$He.  In IA, $H_q^A$ is obtained as a 
convolution between the non-diagonal spectral function of the internal 
nucleons and the GPD $H_q^N$ of the nucleons themselves. 

\begin{figure}[t]
\center
\resizebox{0.45\textwidth}{!}
{\includegraphics{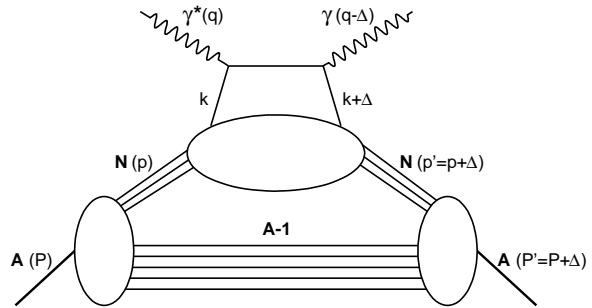}}
\caption{The handbag contribution to the coherent DVCS process
off a nucleus $A$, in IA.}
\label{fig:1}       
\end{figure}

The scenario is depicted in fig.~\ref{fig:1} for the special case of coherent 
DVCS, in the handbag approximation. One parton with momentum $k$, belonging to 
a given nucleon of momentum $p$ in the nucleus, interacts with the probe and 
then reabsorbed, with momentum $k+\Delta$, by the same nucleon, 
without further re-scattering with the recoiling system of momentum $P_R$. 
Finally, the interacting nucleon with momentum $p + \Delta$ is reabsorbed back 
into the nucleus. The IA analysis is quite similar to the usual IA approach to 
DIS off nuclei, the main assumptions being: (i) the nuclear operator is 
approximated by a sum of single nucleon free operators, i.e., there are only 
nucleonic degrees of freedom; (ii) the interaction of the debris originating 
by the struck nucleon with the remnant (A - 1) nuclear system is disregarded, 
as suggested by the kinematics (close to the Bjorken limit) of the processes 
under investigation; (iii) the coupling of the virtual photon with the 
spectator (A-1) system is neglected (given the high momentum transfer), (iv) 
the effect of the boosts is not considered (they can be properly taken into 
account in a light-front framework). It turns out that $H_{q}^A$ can be 
written in the form:
\begin{multline}
H_{q}^A(x,\xi,\Delta^2) = \\ 
\sum_N \int_x^1 \frac{dz}{z}
h_N^A(z, \xi ,\Delta^2 ) 
H_q^N \left( \frac{x}{z},
\frac{\xi}{z},\Delta^2 \right)
\label{main}
\end{multline}
where $\xi=-\Delta^+/2 \bar{P}^+$ and $\Delta^2=(p-p')^2$ are 
the skewness and the momentum transfer to the 
hadron, respectively, $\bar{P} = {(p + p')}/2$ and
\begin{multline}
h_N^A(z, \xi ,\Delta^2 ) =  
\int d E \int d \vec p \, P_N^A(\vec p, \vec p + \vec \Delta,E) \\
\times \delta \left( z + \xi  - \frac{p^+}{\bar P^+} \right)
\label{hq0}
\end{multline}
is the off-diagonal light-cone momentum distribution of the nucleon $N$ in the 
nucleus $A$. Our definition of the light-cone variables is, given a generic 
four vector $a^\mu$, $a^\pm= (a^0 \pm a^3)/\sqrt{2}$. $P_{N}^A (\vec p, 
\vec p + \vec \Delta, E )$ is the one-body off-diagonal spectral function, 
firstly introduced in~\cite{Scopetta:2004kj}, where it is calculated for the 
$^3$He target. $E=E_{min} + E^*_R$ is the so called removal energy, with 
$E_{min}=| E_{A}| - | E_{A-1}|$ and $E^*_R$ is the excitation energy of the 
nuclear recoiling system.

One should notice that eq.~(\ref{main}) fulfills the general properties of 
GPDs~\cite{gpds_here}, i.e., the forward limit reproduces the standard nuclear 
PDF in IA, the first $x$-moment yields the IA form factor. The polynomiality 
property is fulfilled formally but in any calculation using non-relativistic 
wave functions it is actually valid only at order $O( {p^2 \over m^2} )$. 

By taking the forward limit ($\Delta^2 \rightarrow 0, \xi \rightarrow 0$) of 
eq.~(\ref{main}), one gets the expression which is usually found, for the 
parton distribution $q_A(x)$, in the IA analysis of unpolarized DIS:
\begin{eqnarray}
q_A(x_{Bj}) & = &  H_q^A(x_{Bj},0,0) \nonumber \\
& = & \sum_{N} \int_{x_{Bj}/A}^1 { d \tilde z \over \tilde z}
f_{N}^A(\tilde z) \, q_{N}\left( {x_{Bj} \over \tilde z}\right).
\label{mainf}
\end{eqnarray}
In the latter equation,
\begin{eqnarray}
f_{N}^A(\tilde z) & = & h_{N}^A(\tilde z, 0 ,0) \nonumber \\
& = & \int d E \int d \vec p \, P_{N}^A(\vec p,E) 
\delta\left( \tilde z - { p^+ \over \bar P^+ } \right)
\label{hq0f}
\end{eqnarray}
is the light-cone momentum distribution of the nucleon $N$ in the nucleus, 
$q_N(x_{Bj})= H_q^N( x_{Bj} , 0, 0)$ is the distribution of the quarks of 
flavor $q$ in the nucleon $N$ and $P_N^A(\vec p, E)$ is the one body diagonal 
spectral function.

In a typical IA calculation the light-cone momentum distribution $f_{N}^A(z)$ 
turns out to be strongly peaked around the value $z \simeq 1/A$. To select the 
contribution of the nucleons with large ``plus'' momentum fraction one needs 
therefore to be at $z > 1/A$. Looking at the lower integration limit in 
eq.~(\ref{mainf}), it is clear that, in the DIS case, this occurs at 
$x_{Bj} > 1$, where the cross sections are very small. 
Recent analyzes of inclusive data 
at $x_{Bj} > 1$ have only been able to quantify the 
number of such fast, correlated nucleons, but not to really study their internal 
structure~\cite{Egiyan:2005hs,Subedi:2008zz,Hen:2012yva}.

In the coherent channel of a hard exclusive process one has instead a much 
more structured off-diagonal light-cone momentum distribution. In particular, 
the presence of the independent variable $\xi \simeq {x_{Bj} \over 2-x_{Bj}}$ 
helps in obtaining relevant information on non nucleonic degrees of freedom in 
nuclei. Indeed, $\xi$ represents the difference in ``plus'' momentum fraction 
between the initial and final states of the interacting nucleon; since in 
coherent DVCS the nucleus does not breakup, the probability for such a process 
to take place decreases fast with $\xi$. In~\cite{Cano:2003ju}, for the 
deuteron case, it is estimated that the IA predicts a vanishing cross section 
already for $ x_{Bj} \simeq 0.2$, i.e. for $\xi \simeq 0.1$. By experimentally 
tuning $\xi$ in a coherent DVCS process one could therefore explore at 
relatively low values of $x_{Bj}$ contributions to the GPDs not included in 
IA, i.e., non nucleonic degrees of freedom generating correlations at parton 
level or even other exotic effects contributing to the DIS mechanism. In these 
could reside contributions to the explanation of the nuclear anti-shadowing 
and EMC effects (see section ~\ref{sec:3} for  
further discussion).

\section{Spatial distribution of energy, momentum and forces experienced by 
partons in nuclei}
\label{sec:2}

In this section, we shall discuss how the lowest Mellin moments of GPDs 
provide us with information about the spatial distribution of energy, momentum 
and forces experienced by quarks and gluons inside nuclei. This idea, leading 
to a prediction to be tested experimentally, has been developed initially 
in~\cite{Polyakov:2002yz}. To be specific, let us consider 
a spin-$1/2$ hadronic 
target, e.g. a nucleon. All spin independent equations apply to the spin-$0$ 
targets as well.

The $x$-moments of the GPDs are related to the form factors (ffs) of the 
symmetric energy momentum tensor (EMT), whose nucleon matrix element can be 
para\-me\-trized through three scalar ffs, as follows~\cite{Ji:1996ek}:

\begin{eqnarray}
\nonumber
\langle p'| \hat T_{\mu\nu}^{Q}(0)|p\rangle&=&\bar N(p')\biggl[ M_2^{Q}(t)\ 
\frac{\bar P_\mu\bar P_\nu}{m_N}+ J^{Q}(t)\ 
\frac{i\bar P_{\{\mu}\sigma_{\nu\} \rho}\Delta^\rho}{m_N} \\ 
&+&d^{Q}(t)\  
\frac{1}{5 m_N}\ \left( \Delta_\mu\Delta_\nu-g_{\mu\nu}\Delta^2\right) \\ 
\nonumber & + & \bar c(t)g_{\mu\nu} \biggr]N(p)\, .
\label{EMTffs}
\end{eqnarray}
Here $\hat T_{\mu\nu}^Q=\frac i2\ \bar \psi \gamma_{\{\mu} \nablaboth_{\nu\}}
\psi$ is the quark part of the QCD EMT (the gluon case is analogous) and
the normalization
$\bar N N=2\ m_N$ is assumed. The ffs we are 
interested in, $d^Q(t)$ in eq.~(\ref{EMTffs}), is related to the first 
Mellin moment of the unpolarized GPDs~\cite{Ji:1996ek}:

\begin{eqnarray}
\label{jisr} 
\int_{-1}^1 dx\ x\ H(x,\xi,t)=M_2^Q(t) +\frac 45\ d^Q(t)\ \xi^2\, . 
\end{eqnarray}
Thanks to this relation, $d^Q(t)$ can be studied in hard exclusive processes. 
In particular, $d^Q(t)$ contributes with an $x_{Bj}$ independent term to the 
real part of the DVCS amplitude, which is accessible through the beam charge 
asymmetry \cite{Brodsky:1972vv}. At the same time,  this ff is related to the 
so-called D-term in the parametrization of the GPDs \cite{Polyakov:2002wz}. At 
small $x_{Bj}$ and $t$, to the leading order in $\alpha_s(Q)$, the $x_{Bj}$ 
dependent contribution to the real part of the DVCS amplitude is basically 
given by the ``slice" $H_q(\xi,\xi,t)$ of quark GPD, directly measurable in 
the DVCS beam spin asymmetry. In principle, the ff $d^Q(t)$ can be therefore 
extracted from combined data of DVCS beam spin asymmetry and beam charge 
asymmetry.

In the Breit frame, where $\Delta^0=0$ and $t=\Delta^2=
-\vec \Delta^2$, one can 
introduce the static EMT as follows:
\begin{eqnarray}
T_{\mu\nu}^Q(\vec r,\vec s)=\frac{1}{2 E}\int \frac{d^3\Delta}{(2\pi)^3}\
e^{i\vec{r}\cdot\vec{\Delta}}\ \langle p',S'|\hat T_{\mu\nu}^Q(0)|p,S
\rangle\, .
\label{8}
\end{eqnarray}
$S^\mu$ and $S'^\mu$ 
correspond to the polarization vector $(0,\vec s)$ in the rest frame of the 
nucleon. Various components of $T_{\mu\nu}^Q(\vec r,\vec s)$ can be 
interpreted as spatial distributions (averaged over time) of the quark 
contribution to mechanical characteristics of the nucleon. In particular, 
using eqs.~(\ref{EMTffs}) and (\ref{8}), one can show that $d^Q(t)$ is related 
to the traceless part of $T_{ik}^Q(\vec r,\vec s)$, which characterizes the 
spatial distribution (averaged over time) of shear forces experienced by 
quarks in the nucleon~\cite{Polyakov:2002wz}. Considering the nucleon as a 
continuous medium, $T_{ij}^Q(\vec r)$ describes the force experienced by 
quarks in an infinitesimal volume at distance $\vec r$ from the center of the 
nucleon. In particular, at $t=0$, one obtains:
\begin{eqnarray}
\label{EMTd0}
d^Q(0) =-\frac{m_N}{2}\  \int d^3r\   T_{ij}^Q(\vec r)\ \left(r^i r^j-\frac
13\ \delta^{ij} r^2\right)\, .
\end{eqnarray}

First principles predictions are not possible for $d(t)$. Estimates based on a 
chiral quark soliton model \cite{Petrov:1998kf} yield, at a low normalization 
point, $\mu\approx 0.6$~GeV, a rather large and negative value of $d^Q(0) 
\approx -4.0$~\cite{Kivel:2000fg}. The negative sign has a deep relation to 
the spontaneous breaking of chiral symmetry in QCD (see, 
e.g.,~\cite{Goeke:2001tz}).

\begin{figure*}[t]
\center
\resizebox{0.8\textwidth}{!}{\includegraphics{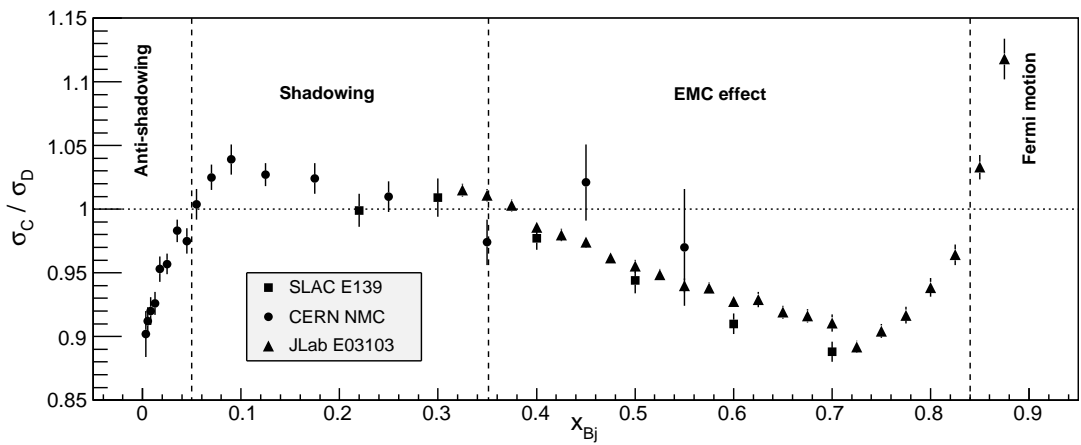}}
\caption{Cross section ratio of lepton scattering on carbon over deuterium in 
the deep inelastic regime from the SLAC E139~\cite{Gomez:1993ri}, CERN 
NMC~\cite{Amaudruz:1995tq} and JLab E03103~\cite{Seely:2009gt} experiments.}
\label{fig:EMC}
\end{figure*}

In ref.~\cite{Polyakov:2002wz}, to illustrate the physics of $d^Q(t)$, a 
simple model of a large nucleus is considered. Generically, for homogeneous 
spin-0 and spin-1/2 targets, one can write:
\begin{eqnarray}
\label{tijdecomp}
T_{ij}(\vec r)=s(r) 
\left(\frac{r_ir_j}{r^2}-\frac 13\ \delta_{ij}\right)+p(r)\delta_{ij}\, .
\end{eqnarray}
The functions $s(r)$ and $p(r)$ are related to each other by conservation of 
the EMT. The function $p(r) $ can be interpreted as the radial distribution of 
the ``pressure" inside the hadron. The function $s(r)$ is related to the 
distribution of the shear forces and, in the model under scrutiny, to the 
surface tension. In fact, one can assume initially that the pressure $p(r)$ 
follows basically the trend of the charge density $\rho(r)$, i.e., it has a 
constant value, $p_0$, in the bulk of the nucleus, and it changes only in the 
thin ``skin" around the radius $R$ of the nucleus. The measurements of 
coherent hard exclusive processes (like DVCS) on nuclei can give detailed 
information about deviations of the energy, pressure, and shear forces 
distributions from that of electric charge. As an illustration, one can 
consider a liquid drop model for a nucleus, with sharp edges. In this case, 
the pressure can be written as
\begin{eqnarray}
p(r)=p_0\ \theta(R-r)-\frac{p_0 R}{3} \delta(R-r)\, .
\end{eqnarray}
Using the condition $\partial_k T_{kl}(\vec r)=0$ in eq.~(\ref{tijdecomp}), 
one obtains
\begin{eqnarray}
\label{sdel}
s(r)=\frac{p_0 R}{2} \delta(R-r)=\gamma\ \delta(R-r)\, ,
\end{eqnarray}
with $\gamma=\frac{p_0 R}{2}$ being the surface tension. Substituting the 
solution (\ref{sdel}) into eq.~(\ref{EMTd0}), $d(0)$ gets the following 
negative value:
\begin{eqnarray}
\label{d0estimate}
d(0)=-\frac{4\pi}{3} m_A\ \gamma\ R^4\, .
\end{eqnarray}

The effect of the finite width of the nuclear ``skin" also has a negative sign 
and the corresponding formula is given in~\cite{Polyakov:2002wz}. Assuming 
that the surface tension depends slowly on the atomic number, as it is 
suggested by nuclear phenomenology, one gets $d(0)\sim A^{7/3}$, i.e. it 
rapidly grows with the atomic number. This fact implies that the contribution 
of the D-term to the real part of the DVCS amplitude grows with the atomic 
number as $A^{4/3}$. This should be compared to the behavior of the amplitude 
$\sim A$ in IA and experimentally checked by measuring the charge beam 
asymmetry in coherent DVCS on nuclear targets. A similar $A$ dependence of 
$d(0)$ has been predicted also in a microscopic evaluation of nuclear GPDs for 
spin-0 nuclei in the framework of the Walecka model~\cite{Guzey:2005ba}. The 
meson (non-nucleonic) degrees of freedom were found to strongly influence DVCS 
nuclear observables, in the HERMES kinematics, at variance with the proton 
case.

The first experimental study of DVCS on nuclei of noble gases, reported 
in~\cite{Airapetian:2009cga}, was not able to observe the predicted $A$ 
dependence. The data are anyway affected by sizable error bars and more 
precise experiments could provide information on nuclear modifications of the 
EMT ffs. The idea in~\cite{Polyakov:2002yz}, summarized here above, has been 
recently retaken in refs.~\cite{Kim:2012ts,Jung:2014jja}, where the 
EMT ffs of the nucleon in nuclear matter have been investigated in different 
effective models of the nucleon structure, i.e., in-medium modified SU(2) 
Skyrme model and $\pi-\rho-\omega$ soliton model, respectively, leading in 
both cases to specific medium effects which could be observed in future DVCS 
experiments off nuclear targets.

\section{Nuclear GPDs and modified nucleon structure}
\label{sec:3}

The study of Nuclear GPDs 
will shed a new light on several longstanding questions about the 
partonic structure of nuclei. In particular, one can wonder how 
the medium modifications of the parton structure of bound nucleons,
observed in DIS and responsible of
the EMC, anti-shadowing and shadowing effects (see fig.~\ref{fig:EMC}),
will be reflected in three dimensional observables such as the \sloppy GPDs. 
These effects are describing the variation of the
nuclear structure functions with respect to the one of 
the deuteron, described by the ratio $R = 2 F_2^A/(A F_2^d)$. 
The shadowing effect is associated with the reduction 
of $R$ at $x_{Bj} < 0.05$, 
the EMC effect with the reduction of $R$ for $0.35 < x_{Bj} < 0.7$ and the 
anti-shadowing with the slight enhancement between 
them. The EMC effect is usually described as a modification of the partonic 
content of nuclei, either linked to an alteration of the nucleons composing 
them or to the addition of non nucleonic components. As we will see, these 
assumptions lead to very different predictions for the nuclear GPDs. The 
shadowing region is usually associated with coherent effects due to the 
interaction length larger than the internucleon separation in nuclei 
(see, e.g., \cite{Piller:1999wx}). In which case, the cross section is 
governed by the surface seen by the photon and behaves like $A^{2/3}$
instead of $A$. This hypothesis can be adapted to nuclear DVCS and tested 
against DVCS' observables.

\subsection{The EMC region}

\begin{figure}[t]
\center
\resizebox{0.45\textwidth}{!}
{\includegraphics{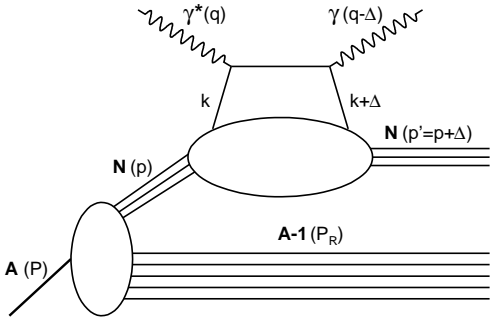}}
\caption{The handbag contribution to the incoherent DVCS process
off a nucleus $A$, in IA.}
\label{fig:3}       
\end{figure}

DVCS on nuclei can occur through two mechanisms, namely
coherent DVCS, shown in fig. \ref{fig:1}, which gives access to the GPDs of 
the nucleus as a whole, and incoherent DVCS, shown in fig. \ref{fig:3}, 
which gives access to in-medium nucleon GPDs. 
The measurement of nuclear GPDs will allow to 
localize the partons in the transverse plane providing,
in the valence region, a pictorial description of
the EMC effect observed in DIS. In the 
case of incoherent DVCS, the comparison between free and in-medium 
nucleons allow to explore the variation with $t$ of 
a properly defined generalization of the EMC ratio,
providing the usual one in the forward limit, 
at $t=0$.

Nuclei of spin-0 ($^4$He, $^{12}$C, $^{16}$O...) are especially good 
candidates for these studies because of their simplicity, indeed at leading 
twist they are described by a single chiral-even GPD $H(x,\xi,t)$%
\footnote{If we do not neglect the mass of quarks, a single chiral-odd GPD 
($H_T(x,\xi,t)$) also contributes to the structure of the nuclei.}.
In general, GPDs are not observables. In the DVCS amplitude
they appear in the so called Compton Form Factors (CFFs),
convolution integrals in the non-observable $x$ variable. 
CFFs are observable quantities, depending on the experimental
variables $\xi$ and $t$.
Both the real and imaginary parts of the CFF associated to
the GPD $H(x,\xi,t)$
can be uniquely extracted from DVCS 
beam spin asymmetry and beam charge asymmetry
using their different $sin \phi$ 
and $cos \phi$ contributions to the cross section, where
$\phi$ is the azimuthal angle of the detected photon
with respect to the leptonic plane
(see~\cite{Kirchner:2003wt,%
Belitsky:2008bz} for exact formulas). 

In order to describe, in the GPD framework,
the nucleon medium modifications, leading to
the EMC effect in the inclusive limit, 
three ways have 
been followed: (i) Liuti {\it et al.} 
have given a description including dynamical off-shellness of the nucleons
~\cite{Liuti:2005gi,Liuti:2005qj,Liuti:2004hd},
i.e., allowing for medium modification of the nucleon parton
structure beyond the conventional binding and Fermi motion ones, already
included in the spectral function used in IA analyzes; (ii)
Guzey and Siddikov~%
\cite{Guzey:2005ba} have  
included meson degrees of freedom~\cite{Guzey:2005ba}; 
(iii) finally, in another report,
medium modified form factors have been included by 
Guzey {\it et al.}~\cite{Guzey:2008fe,Guzey:2009pv}. 

\begin{figure}[t]
\center
\resizebox{0.4\textwidth}{!}{\includegraphics{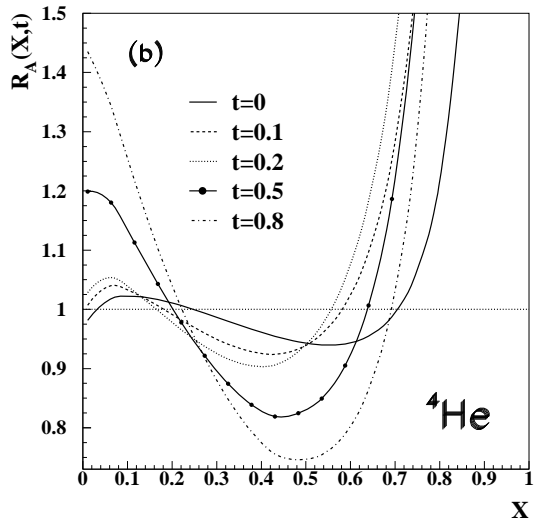}} 
\caption{
Predictions from Liuti {\it et al.}~\cite{Liuti:2005qj} 
for the ratio eq. (\ref{ratliu}).}
\label{fig:LiutiEMC}
\end{figure}

The work of Liuti {\it et al.}~\cite{Liuti:2005gi,Liuti:2005qj} 
includes both a realistic 
nuclear spectral function, leading to conventional nuclear effects
and kinematical off-shellness,
and dynamical off-shellness:
\begin{multline}
H^A_q(x,\xi,t) =   \\
\int \frac{d^4 P}{(2\pi)^4}
H^{N_{OFF}}_{q}(x_N,\xi_N,P^2,t) \, {\cal M}^A({P},{P_{A}},\Delta), 
\label{convo1}
\end{multline}
where ${\cal M}^A$ is the nuclear matrix element and the nucleon is off its 
mass shell ($P^2 \neq M^2$), a feature affecting directly
the nucleon GPD. The latter effect is found 
to be strongly linked to transverse degrees of freedom and therefore 
leads to a strong variation of the structure function with $t$,
at zero skewdness.
This is seen in fig.~\ref{fig:LiutiEMC}, where the ratio
\begin{equation}
R_A(x,\xi=0,t) = {H_A(x,\xi=0,t) F_N(t) \over H_N(x,\xi=0,t) F_A(t)}
\label{ratliu}
\end{equation}
is shown. In the figure, the curve is plotted as a function of
the asymmetric momentum fraction $X$ 
(see, e.g., \cite{GolecBiernat:1998ja})
and not as a function of the standard $x$, but at zero skewdness they have the same value.
Traditional Fermi motion and binding effects do not show such behavior, making 
this observation a direct test of the importance of off-shellness to explain 
the EMC effect. Liuti {\it et al.} also 
consider the long range effects and the 
coupling of the virtual photon to mesons and resonances in nuclei, but 
conclude that none of these mechanisms
contribute significantly to nuclear GPDs. 

Guzey and Siddikov~\cite{Guzey:2005ba} have however very different findings 
when including mesons in nuclei. They use the IA taking into account
meson degrees of freedom, i.e. an expression for the nuclear GPD
which is an extension of eq.~(\ref{main})
\begin{equation}
H^A_{q}(x,\xi,t)=\sum_i\int_{x}^1 \frac{dz}{z} H^A_{i}(z,\xi,t)
H^i_{q}\left(\frac{x}{z},\frac{\xi}{z},t\right)~,
\label{ConvolutionDefinition2}
\end{equation}
where $H^A_{i}$ is the distribution of the hadronic constituents in the 
nucleus (nucleons and mesons) based on the Walecka model~\cite{Serot:1997xg} 
and $H^i_{q}$ is the distribution of the quarks in these hadrons. In the 
latter function, no dynamical off shell-effects are included. They find that 
the meson contribution has a very strong impact, enhancing the charge 
asymmetry and suppressing the spin asymmetry for large $A$, as shown in 
table~\ref{tabasym}.

\begin{table}[t]
\center
\begin{tabular}{|l|c|c|}
\hline
Nucleus & $A^{cos}_{C\,\,A}/A^{cos}_{C\,\,N}$ & 
$A^{sin}_{LU\,\,A}/A^{sin}_{LU\,\,N}$ \\
\hline
\rule{0pt}{2.5ex}%
$ ^{12}C   $  & 4.61  & 2.49 \\
$ ^{16}O   $  & 5.41  & 2.33 \\
$ ^{40}Ca  $  & 7.34  & 1.60 \\
$ ^{90}Zr  $  & 6.80  & 0.81 \\
$ ^{208}Pb $  & 6.12  & 0.31 \\
\hline
\end{tabular}
\caption{\label{tabasym} The predictions for ratios of the nuclear to the free 
proton asymmetries from Guzey and Siddikov~\cite{Guzey:2005ba}.}
\end{table}

\begin{figure}[t]
\center
\resizebox{0.4\textwidth}{!}{\includegraphics{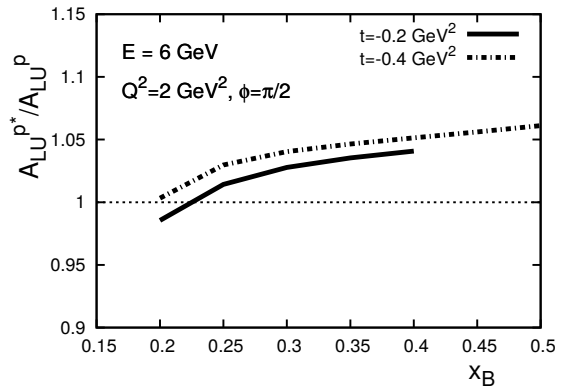}}
\caption{Predictions from Guzey {\it et al.}~\cite{Guzey:2008fe} for the ratio 
of bound to free beam spin asymmetry.}
\label{fig:GTT_Ratio}
\end{figure}

Finally, Guzey {\it et al.}~\cite{Guzey:2008fe} have explored the possibility 
to apply medium modification to the GPDs in a similar way than  
medium modified form factors:
\begin{eqnarray}
\nonumber
H_{q}^{p^{\ast}}(x,\xi,t,Q^2)&=&\frac{F_1^{p^{\ast}}(t)}{F_1^p(t)} \,
H_{q}^p(x,\xi,t,Q^2) \,, \\
E_{q}^{p^{\ast}}(x,\xi,t,Q^2)&=&\frac{F_2^{p^{\ast}}(t)}{F_2^p(t)} 
E_{q}^p(x,\xi,t,Q^2) \,, \\ \nonumber
\tilde{H}_{q}^{p^{\ast}}(x,\xi,t,Q^2)&=&\frac{G_1^{\ast}(t)}{G_1(t)} \,
\tilde{H}_{q}^p(x,\xi,t,Q^2) \,,
\label{eq:gpds_mm}
\end{eqnarray}
where $F_1^{p}$, $F_2^{p}$ and $G_1$ are respectively the Dirac, Pauli and 
axial form factors of the proton. The starred items refer to the bound 
proton calculated using the quark meson coupling model~\cite{Lu:1998tn}. 
This description of the bound nucleons gives rise to an effect opposite 
to the one predicted by Liuti {\it et al.}, 
i.e., a  ratio $A_{LU}^{p^\ast}/A_{LU}^{p}$ which
grows with $x_{B}$ as can be seen in fig.~\ref{fig:GTT_Ratio}.

\subsection{The shadowing region and gluon GPDs}

In the low $x_{Bj}$ region, the contribution of gluons is very important and 
is especially interesting in the case of nuclei. Indeed, saturation is 
expected to impact the gluon distribution in nuclei at higher $x_{Bj}$ with
respect to what happens for the free nucleon~\cite{Piller:1999wx}. 
Moreover, gluons in nuclei are poorly known and 
it is unclear how the nuclear effects observed for quarks (EMC, anti-shadowing 
and shadowing) affect the gluons. 

\begin{figure}[b]
\center
\resizebox{0.4\textwidth}{!}{\includegraphics{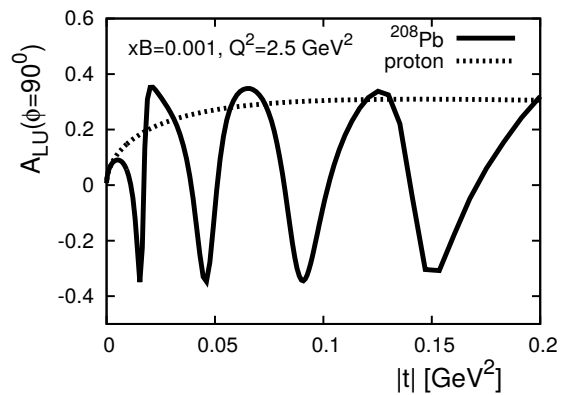}}
\caption{Predictions from~\cite{Goeke:2009tu} for the coherent 
nuclear beam spin asymmetry.}
\label{fig:GGS}
\end{figure}

\begin{figure}[t]
\center
\resizebox{0.3\textwidth}{!}{\includegraphics{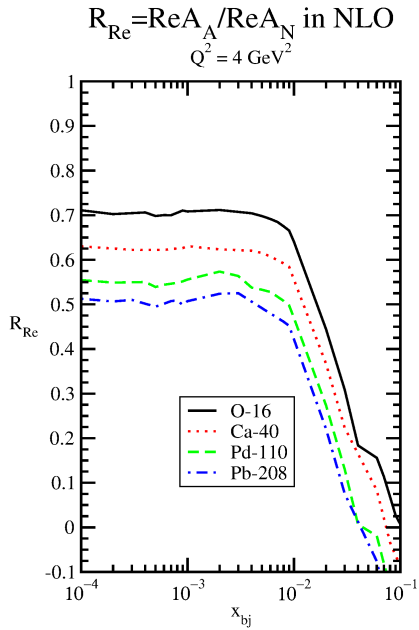}}
\caption{Predictions from~\cite{Freund:2003wm} for the nuclei to nucleon ratio 
of the real parts of the $H$ CFF.}
\label{fig:RealH-EMC}
\end{figure}

By studying shadowing on the $H$ GPD in spin-0 nuclei at low 
$x_{Bj}$, several authors have predicted a stronger effect on GPDs 
than on PDFs~\cite{Freund:2003ix,Freund:2003wm,Goeke:2009tu}. 
They pointed out the high sensitivity of their result to the 
gluon distributions as well. The imaginary part of the CFF related to the GPD
$H$ is indeed predicted to experience a stronger shadowing and to be largely 
affected by the gluon distribution at $x_{Bj}$ as high as 0.1. This leads to 
a very original effect, the oscillation of $A_{LU}$ as a function of $t$,
predicted in~\cite{Goeke:2009tu} and shown in fig.~\ref{fig:GGS}. The 
real part of the CFF related to the GPD
$H$ is also predicted to be strongly affected 
in  spin-0 nuclei with a strong suppression in the $0.01 < x_{Bj} < 0.1$ 
range, seen in fig.~\ref{fig:RealH-EMC}. This effect is due to the 
cancellation of the ERBL and DGLAP region contributions to the real part 
of the CFF related to the GPD $H$ (see~\cite{Goeke:2009tu} 
for a detailed discussion).

\begin{figure}[b]
\center
\resizebox{0.45\textwidth}{!}{\includegraphics{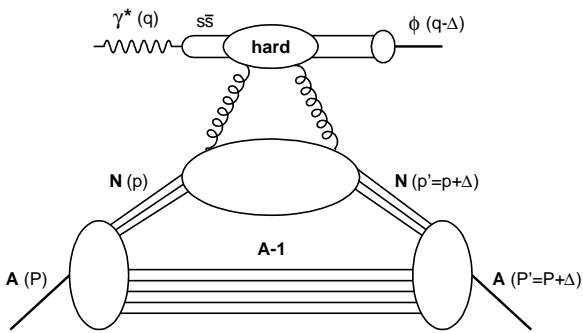}}
\caption{Feynman diagram of the hard production of a $\phi$ meson
illustrating its link to gluon GPDs.}
\label{fig:PhiProd}
\end{figure}

Since, at leading order, gluons do not couple to the photons, they cannot be 
accessed directly with the DVCS process. Deep virtual meson production 
(DVMP) can be a perfect tool to measure the gluon GPDs. This is especially 
true for $\phi$ meson production because of the dominance of the $s\bar s$ 
component that suppresses quark exchange channels and enhances the gluon 
contribution (fig.~\ref{fig:PhiProd}). Therefore, when producing the $\phi$ 
meson, we effectively probe the gluon structure of the target. Work from~%
\cite{Goloskokov:2007nt} shows how one can extract the gluon GPDs of a proton 
target using exclusive $\phi$ lepto-production. The extension to nuclei is not 
so straightforward, in particular at intermediate energies, where it was 
suggested that factorization might not hold~\cite{Rogers:2005bt}. However, the 
uniqueness of this probe into the gluon content of nuclei deserves further 
theoretical work to be done in order to analyze possible future data from JLab.

\subsection{Experimental perspectives}

\begin{figure}[t]
\center
\resizebox{0.4\textwidth}{!}{\includegraphics{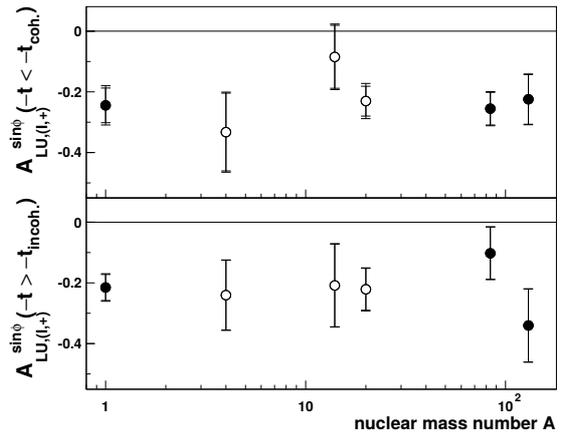}}
\caption{Results of the HERMES collaboration \cite{Airapetian:2009cga} 
for the sinus moment of the  beam spin asymmetry as a 
function of the mass number $A$ for coherent (top) and incoherent (bottom) 
enriched data samples.}
\label{fig:HERMES}
\end{figure}

The first experiment to explore GPDs for nuclei
has been performed by the HERMES collaboration in DESY~%
\cite{Airapetian:2009cga}. However, they 
could not differentiate coherent and incoherent channels directly 
and had to rely on the dominance of either channel at small and large $t$,
respectively. They found no modification of the asymmetries with $A$ in either $t$ sectors (cf.
fig.~\ref{fig:HERMES}), while a basic description of the nuclei 
in terms of the constituent nucleons~\cite{Kirchner:2003wt,%
Guzey:2003jh} predicts an important combinatorial enhancement of asymmetries in 
the coherent region. However, the difficulty to decipher the coherent and 
incoherent channels in HERMES data makes it difficult to reach a strong 
conclusion.

As the HERMES results~\cite{Airapetian:2009cga} have shown, the measurement of 
nuclear DVCS is very difficult. This difficulty lies in the large energy gap 
between the high energy photons and the slow recoiling nuclei, which need 
very different detector systems to be measured in coincidence. The CLAS 
collaboration at JLab has performed a measurement of coherent DVCS on $^4$He 
which is still under analysis. The preliminary results indicate that they were 
successful in measuring both coherent and incoherent DVCS channels exclusively~%
\cite{Voutier:2013gia,hattawy}. While these results are not released yet, the 
preliminary analysis clearly shows only a small coverage in $x_{Bj}$ and $t$ 
and we should expect that an extension of this program with the upgraded CLAS12 
will provide a large data set to analyze light nuclei GPDs in the valence 
region. Farther in the future, the project of an electron-ion collider in the 
US~\cite{Accardi:2012qut} will be the perfect tool to study nuclear DVCS. 
Indeed, because of the collider kinematics, it will be much easier to 
detect the recoiling nuclei and to polarize the incoming nuclei. 
Together with the high energy available, the electron ion collider 
will allow to cleanly map the nuclear GPDs at low $x$, including gluon GPDs.  

\section{Flavor separation using light nuclei}

Since conventional nuclear effects, if not properly evaluated, can be easily 
mistaken for exotic ones, light nuclei, for which realistic calculations are 
possible and conventional nuclear effects can be calculated exactly, play a 
special role. Besides, light nuclei impose their relevance in 
the extraction of the neutron information, necessary to perform a clean flavor 
separation of GPDs and TMDs, crucial to test QCD fundamental symmetries and 
predictions. We note that an indirect procedure to constrain the neutron GPDs 
using coherent and incoherent DVCS off nuclei has been proposed 
in~\cite{Guzey:2008th}.

In the following two subsections, the help one can get from studies of light 
nuclei will be summarized for GPDs and TMDs, respectively, in particular for 
the $^3$He target.

\subsection{GPDs}

\begin{figure}[t]
\center
\vspace{-3cm}
\hspace*{-2cm}
\resizebox{0.7\textwidth}{!}{\includegraphics{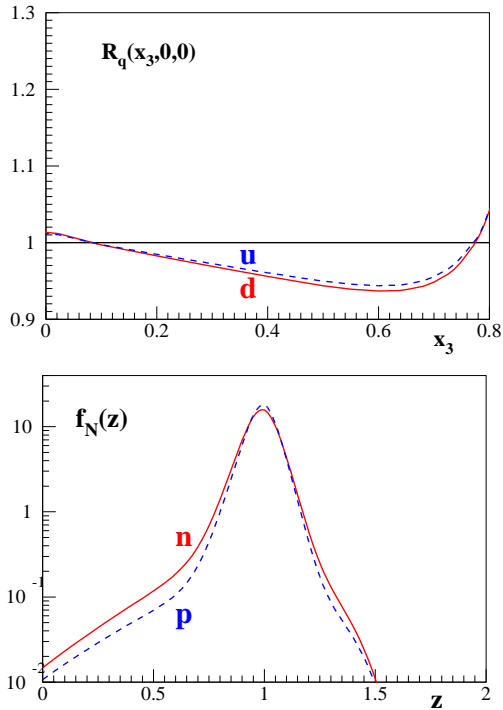}}
\vspace{-5cm} 
\caption{Upper panel: the dashed (full) line represents the ratio of the 
$^3$He GPD $H$ to the corresponding quantity of the constituent nucleons (2 
protons and one neutron), for the u (d) flavor, in the forward limit, as a 
function of $x_3=3x$. Lower panel: the dashed (full) line represents the light 
cone momentum distribution, eq.~(\ref{hq0f}), for the proton (neutron) in 
$^3$He.}
\label{fig:lcforward}       
\end{figure}

As it has been shown in section 1, the conventional treatment of nuclear GPDs, 
through IA, involves a non-diagonal nuclear spectral function. The complicated 
dependence on the momentum and removal energy of the spectral function can be 
evaluated exactly for $^3$He, which is therefore simple enough to allow a 
realistic treatment and very suitable, being not scalar, for polarization 
studies and, being not isoscalar, for flavor separation. 

A realistic microscopic calculation of the unpolarized quark GPD of the $^3$He 
nucleus has been presented in~\cite{Scopetta:2004kj}. The proposed scheme 
points to the coherent channel of hard exclusive processes. Nuclear effects, 
evaluated within the AV18 potential~\cite{Wiringa:1994wb}, are found to be 
larger than in the forward case and increase with increasing $t$ and keeping 
$\xi$ fixed, and with increasing $\xi$ at fixed $t$. Besides, the obtained GPD 
cannot be factorized into a $t$-dependent and a $t$-independent term, as 
suggested in prescriptions proposed for finite nuclei. 

\begin{figure}[t]
\center
\vspace{-3cm}
\hspace*{-2cm}
\resizebox{0.7\textwidth}{!}{\includegraphics{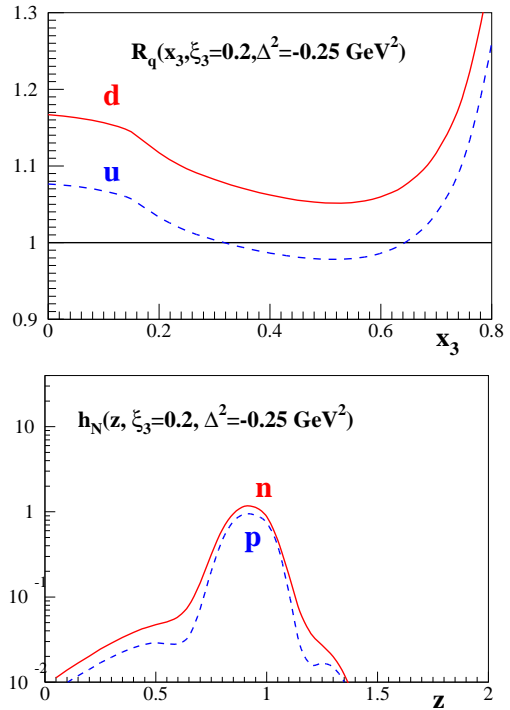}}
\vspace{-5cm}       
\caption{Upper panel: the same as in the upper panel of the previous figure, 
but at off-forward kinematics: $t = -0.25$ GeV$^2$ and $\xi_3 = 3 \xi = 0.2$. 
Lower panel: the dashed (full) line represents the light cone off diagonal 
momentum distribution, eq.~(\ref{hq0}), for the proton (neutron) in $^3$He at 
the same off-forward kinematics.}
\label{fig:lcoff}       
\end{figure}

In~\cite{Scopetta:2009sn}, the analysis has been extended, showing that other 
conventional nuclear effects, such as isospin and binding ones, or the 
uncertainty related to the use of a given nucleon-nucleon potential, are 
rather bigger than in the forward case. An example is seen in 
figs.~\ref{fig:lcforward} and~\ref{fig:lcoff}. Clearly, nuclear effects 
increase when the light-cone momentum distributions, eqs.~(\ref{hq0}) 
and~(\ref{hq0f}), depart from a delta-like behavior. Besides, nuclear effects 
for the $u$ ($d$) flavor follow the path of the proton (neutron) light-cone 
momentum distributions. The experimental check of this behavior, typical 
prediction of a realistic conventional IA approach, which should not show up 
in an isoscalar target, such as $^2$H or $^4$He, would give relevant 
information on the reaction mechanism of DIS off nuclear targets.

In ref.~\cite{Rinaldi:2012pj} the issue of the extraction of the neutron 
information, in particular the one related to the parton angular momentum 
content, accessible in principle through the Ji's sum rule if also the GPD $E$ 
is measured, has been addressed. Whenever properties related to the 
polarization of the neutron have to be studied, $^3$He is an ideal target, 
since at a 90\% level it is equivalent to a polarized neutron. It was found 
that the sum of $H$ and $E$ is dominated to a large extent by the neutron 
contribution. 
A technique has been therefore proposed~%
\cite{Rinaldi:2012ft}, able to take into account the nuclear 
effects included in the IA analysis and to safely extract the neutron 
information at values of the momentum transfer large enough to allow the 
measurements. 
A similar extraction technique has been successfully tested for the extraction 
of the $\tilde H$ GPD from the corresponding quantity of $^3$He 
in~\cite{Rinaldi:2014bba}. In this case, this investigation would require 
coherent DVCS off polarized $^3$He, a challenging but not impossible 
measurement at present facilities~\cite{Charles:2015}. 
Thanks to this observations, coherent DVCS should be considered a 
key experiment to access the neutron GPDs and, in turn, the orbital angular 
momentum of the partons in the neutron. One should notice that isoscalar 
targets, such as $^2$H and $^4$He, have a very small contribution from the 
$E$ GPD and are not useful for this investigation. The measurement of the $E$ 
GPD would require anyway transverse polarization of $^3$He and a very 
difficult measurement in the coherent channel, at the present facilities. The 
other way to obtain the neutron information could be through incoherent DVCS 
off the deuteron, a process which is hindered by FSI; 
specific kinematical regions, where FSI are known
to be less relevant, have to be therefore selected and
dedicated theoretical estimates of FSI in this channel
will be very important. An experiment
of this kind has been approved at JLab and will run
after the 12 GeV upgrade~\cite{silvia}.
Another promising possibility for the measurement
of DVCS off the neutron, to be detailed in forth-coming
proposals~\cite{Hafidi:2009loi}, is that offered by the detection
of a slow recoiling proton in DVCS off the deuteron, 
exploiting the experimental setup successfully
used in spectator SIDIS by the BONUS
collaboration at JLab~\cite{KKW}. We note in passing that,
for the deuteron target, the coherent channel
has been thoroughly studied theoretically 
\cite{Berger:2001zb,Cano:2003ju}, showing that
coherent measurements are possible and
would be very interesting. However, 
to fully unveil the rich GPDs structure of this spin-1 system,
one should be able to polarize the target, a rather
complicated issue at present.

In this scenario, 
$^3$He represents an important target for nuclear GPDs studies. 
Its conventional structure is completely under control, and it is ideal to 
check the interplay of conventional and exotic effect, as a playground to have 
hints on them when heavier nuclear targets are used. Besides, it is a unique 
effective polarized neutron and the neutron $E$ and $\tilde H$ GPDs at low $t$ 
could be extracted easily from the corresponding $^3$He quantities, with 
little model dependence. This would require measurements of coherent DVCS, 
certainly challenging but, for $\tilde H$, unique and not prohibitive.

\subsection{TMDs}

The most natural process to obtain information on the 3D nucleon structure in 
momentum space is SIDIS, i.e. the process where, besides the scattered lepton, 
a hadron is detected in coincidence.  If the hadron is fast, one can expect 
that it originates from the fragmentation of the active, highly off-mass-shell 
quark, after absorbing the virtual photon. Hence, the  detected hadron carries 
valuable information about the motion of quarks in the parent nucleon before 
interacting with the photon, and in particular on their transverse motion. 
Therefore, through SIDIS reactions, one can access TMDs (see, e.g., 
refs.~\cite{gpds_here,Barone:2001sp,Barone:2010zz}) and try to shed some light 
on issues which cannot be explained in the collinear case, such as the 
phenomenology of the transversity PDF, the solution to the spin crisis and, in 
the nuclear case, the mechanism of nuclear DIS processes and the EMC effect. 
In order to experimentally investigate the  wide field of TMDs, one should 
measure cross-section asymmetries, using different combinations of beam and 
target polarizations (see, e.g., ref.~\cite{D'Alesio:2007jt}). In particular, 
single spin asymmetries (SSAs) with transversely polarized targets 
$\overrightarrow A$ allow one to experimentally distinguish the Sivers and the 
Collins contributions, expressed in terms of different TMDs and fragmentation 
functions (FFs)~\cite{Barone:2001sp}. A large Sivers asymmetry was measured in 
${{{\overrightarrow p}(e,e'\pi)x}}$~\cite{Airapetian:2004tw} and a small one 
in ${{{\overrightarrow D}(e,e'\pi)x}}$~\cite{Alexakhin:2005iw}, showing a 
strong flavor dependence of TMDs. To clarify this issue, high precision 
experiments involving both protons and neutrons are needed. This puzzle has 
attracted a great interest in obtaining new information on the neutron TMDs.

The possibility to extract information on neutron \linebreak TMDs from 
measurements of the SSAs in the processes \linebreak $\overrightarrow{^3He}
(e,e'\pi^\pm)$, using transversely polarized targets, was \linebreak used in a 
series of experiments at JLab Hall-A~\cite{Qian:2011py,Allada:2013nsw}, and it 
will be used again after the 12~GeV upgrade~\cite{Cates}.

We have seen that polarized $^3$He is an ideal target to study the neutron 
spin structure. To obtain a reliable information one has to take carefully 
into account: (i) the nuclear structure of  $^3$He, (ii) the interaction in 
the final state (FSI) between the observed pion and the remnant debris, and 
(iii) the relativistic effects. 

Dynamical nuclear effects in inclusive deep inelastic electron scattering 
{{$^3\overrightarrow{He}( e,e')X$}} (DIS) were evaluated~%
\cite{CiofidegliAtti:1993zs} with a realistic $^3{{\overrightarrow{He}}}$ 
spin dependent spectral function
It was found that the formula
\begin{equation}
{{A_n }}\simeq {1 \over {{p_n}} f_n} 
\left ( {A^{exp}_3} - 2 {p_p} f_p ~ {{A^{exp}_p}} \right ) 
\label{formula}
\end{equation}
can be safely adopted to extract the neutron information, the asymmetry $A_n$, 
from the corresponding quantities for the proton and $^3$He. This formula is 
actually widely used by experimental collaborations (see, e.g. 
ref.~\cite{Abe:1997cx}). The nuclear effects are hidden in the proton and 
neutron {{"effective polarizations"}} (EPs), $ p_{p(n)}$. 
$f_{p(n)}$ in eq.~(\ref{formula}) are the dilution factors.

\begin{figure}[t]
\center
\resizebox{0.45\textwidth}{!}{\includegraphics{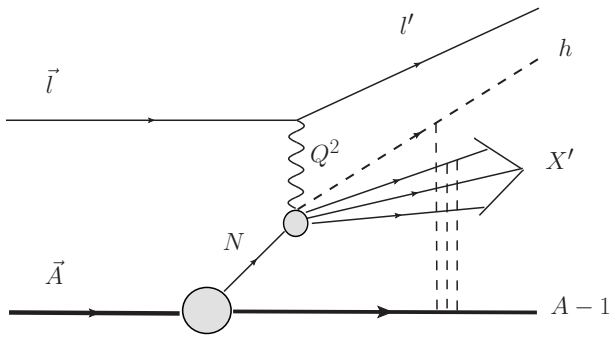}}
\caption{Interaction between the $(A-1)$ spectator system
and the debris produced by the absorption of a virtual photon by
a nucleon in the nucleus.}
\label{gea}     
\end{figure}

To investigate if an analogous formula can be used to extract the SSAs, 
in~\cite{Scopetta:2006ww} the processes {{$^3\overrightarrow{He}
(e,e'\pi^{\pm})X$}} were evaluated in the Bjorken limit and in IA.
In such a framework, {{SSAs}} for $^3He$ involve 
convolutions of the spin dependent spectral function with 
TMDs and FFs. Ingredients of the calculations were: (i) a realistic 
spin dependent spectral function,
obtained using the {{AV18}} 
interaction~\cite{Wiringa:1994wb}
; (ii) parametrizations of data or models for TMDs and FFs; 
The extraction procedure 
through the formula successful in DIS was found to work nicely for both the 
Sivers and Collins SSA. The generalization of eq.~(\ref{formula}) to extract 
the neutron information was recently used by experimental collaborations~%
\cite{Qian:2011py,Allada:2013nsw}. The question whether FSI effects can be 
neglected was anyway a missing point in the analysis of~%
\cite{Scopetta:2006ww}. This problem has been faced in~\cite{Kaptari:2013dma}. 
In SIDIS experiments off $^3$He, the relative energy between the spectator 
$(A-1)$ system and the system composed by the detected pion and the remnant 
debris (see fig.~\ref{gea}) is a few GeV and FSI can be treated through a generalized 
eikonal approximation (GEA). 

\begin{figure}
\center
\resizebox{0.48\textwidth}{!}{\includegraphics{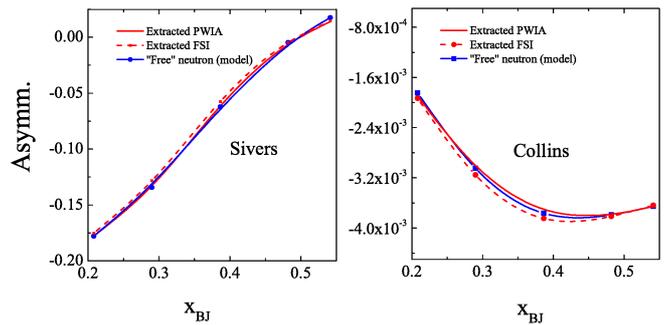}}
\caption{Check of the extraction procedure, eq.~(\ref{formula}), with
and without FSI taken into account, for the Sivers (left) and Collins (right)
SSAs, in the kinematics of \cite{Cates}.}
\label{3hetmd}     
\end{figure}

The GEA was already successfully applied to 
nicely describe data of unpolarized spectator SIDIS off the deuteron~%
\cite{KKW} in ref.~\cite{Atti:2010yf}. The FSI effects to be considered 
are due to the propagation of the debris, formed after the $\gamma^*$ 
absorption by a target quark, and the subsequent hadronization, both of them 
influenced by the presence of a fully interacting $(A-1)$ spectator system 
(see fig.~\ref{gea}). Within the GEA, the key quantity to introduce FSI is the 
{\em{distorted}} spin dependent spectral function, a complicated object 
defined through overlaps between the $^3$He wave function and that of the 
particles in the final state, fully interacting through Glauber 
re-scatterings. The model parameters can 
be found in~\cite{CiofidegliAtti:2003pb}. As a consequence of 
FSI, from the IA calculation to the GEA one, in the kinematics 
of \cite{Cates}, the EPs change considerably. Anyway, one has to 
consider also the effect of the FSI on dilution factors.
It was found, in a wide 
range of kinematics, typical for the experiments at JLab~\cite{Cates}, 
that the product of EPs and dilution factors changes very little~%
\cite{DelDotto:2014zua}, the effects of FSI in the dilution factors and in 
the EPs compensate each other to a large extent and the usual extraction,
given in eq. (\ref{formula}), appears to be safe.
Therefore, nuclear effects driven by the GEA description of FSI are safely 
taken care of by the simple extraction formula eq.~(\ref{formula}) (see 
fig.~\ref{3hetmd}). Relativistic effects are under consideration and 
preliminary results have been presented in~\cite{DelDotto:2014zua}.

\section{Nuclear transverse momentum dependent parton distributions}

As we have seen in the previous section, SIDIS cross sections and the related 
azimuthal asymmetries can be expressed in terms of TMDs. This is an important 
focus in recent  studies of the nucleon structure~\cite{gpds_here}, and, in 
principle, one could use the same framework to study nuclei, although 
calculations involving many nucleons can be tedious. Z.T.~Liang{\it et al.}~%
\cite{Liang:2008vz,Gao:2010mj,Song:2013sja,Song:2014sja} have shown how higher 
twist nuclear effects on TMDs can be simply expressed in term of a transport 
parameter, typical of cold nuclear matter:
\begin{equation}
f_q^A(x,k_\perp)\approx\frac{A}{\pi \Delta_{2F}} \int d^2\ell_\perp 
e^{-(\vec k_\perp -\vec \ell_\perp)^2/\Delta_{2F}}f_q^N(x,\ell_\perp).
\label{eq:tmdNA}
\end{equation}
where $\Delta_{2F}$ is the average local transport parameter 
experienced by the struck quark on its path through the nuclear medium. 
\begin{equation}
\Delta_{2F}=\int d\xi^-_N \hat q_F(\xi_N).
\label{eq:Delta2F}
\end{equation}
The local transport coefficient $\hat q_F(\xi_N)$ of the nuclear medium is 
defined as the mean transverse momentum squared it induces on a fast parton 
going through it, $\xi_N$ being the position in the nucleus in light-cone 
coordinates.  It can be indirectly accessed in many hadronization processes, 
in which it leads
to transverse momentum broadening or jet broadening~\cite{Accardi:2009qv}. 
Such experiments
have suggested values of $\hat q$ ranging from 0.075 to 
0.75~GeV$^2$/fm in cold nuclear matter~\cite{dupre}. The 
possibility to measure $\hat q$ through TMDs would give an essential 
cross check
on the highly model dependent extraction of this fundamental nuclear parameter.
Indeed, the quark transport parameter in nuclei is directly linked to 
the gluon distribution at $x \rightarrow 0$ \cite{Baier:1996sk}:
\begin{equation}
\hat q_F(\xi_N)
=\frac{2\pi^2\alpha_s}{N_c}\rho_N^A(\xi_N)[xf^N_g(x)]_{x \rightarrow 0},
\label{qhat1}
\end{equation}
where $\rho_N^A(\xi_N)$ is the local nucleon density in the nucleus and 
$f^N_g(x)$ is the gluon distribution function. One can also directly relate 
this observable to the saturation scale, where $f^N_g(x)$ is maximum
(see, e.g., ~\cite{Kopeliovich:2010aa}).

\begin{figure}[t]
\center
\resizebox{0.45\textwidth}{!}{\includegraphics{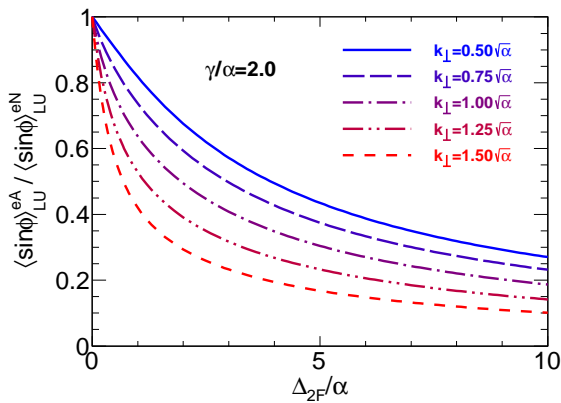}}
\caption{Ratio of the nuclear $\langle \sin \phi \rangle _{LU}$ beam spin 
asymmetry to the nucleon one.}
\label{fig:TMD}
\end{figure}

In their various studies, Z.T.~Liang {\it et al.}, show that the transport 
reduces the azimuthal asymmetries in most of the phase space (see 
fig.~\ref{fig:TMD} for example). A measurement of this effect,
which would give an independent measurement of $\hat q$,
has been proposed at JLab~\cite{Accardi:2014loi}.
Moreover, a precise measurement of the TMD asymmetries 
would hint at possible modifications of the nucleon in-medium, in terms of its
transverse momentum degrees of freedom. However, we are not aware of
any prediction on this last topic.

\section*{Conclusions}

While experimental data are still scarce in the domain, the 3D imaging of 
nuclei has already strong theoretical basis and numerous strong motivations. 
In particular, we highlighted the possibility to isolate non nucleonic 
degrees of freedom in nuclei and the new possibility to measure 
the shear force and pressure distribution in nuclei, 
offered by the GPDs description of hard exclusive processes.

We showed the great hope that can be placed in the GPD framework applied to
nuclei in order to solve the conundrum on the EMC effect and its numerous 
different explanations. Indeed, the 3D imaging of the nuclei 
will allow to locate
where the EMC effect is stronger in the transverse plane. This would offer
some really new data, for which nuclear models offer very different 
predictions and could be distinguished.

From a practical point of view, we have seen that the use of spin-0 
targets simplifies the formalism, allowing for a limited number of 
measurements to make an important impact. Also the use of light nuclei, whose 
internal dynamics is well known in term of nucleons, eases the theoretical
description and is important to allow for a precise flavor separation
of GPDs and TMDs.
Most importantly, it makes possible to detect the intact nuclei 
in actual experiments. As we have seen, the identification of the coherent and 
incoherent channels is very important to interpret the data, which is the 
biggest challenge for future experimental projects.

At the low end of the $x$ spectrum, in the shadowing region, the models
we have reviewed predict very strong nuclear effects for the GPDs and
therefore the DVCS observables. 
The project for an electron ion collider~\cite{Accardi:2012qut} appears
to be the best facility in order to test these predictions. Among them,
the oscillation of the beam spin asymmetry signal with
$t$ at low $x_{Bj}$ seems the most original.

We showed how TMDs can be used to independently measure the nuclear
transport parameter $\hat q$ and how it directly relates to the gluon 
distribution at $x \rightarrow 0$ and to the saturation scale in nuclei.
The extraction of $\hat q$ using hadronization data has lead to very
different results and is highly model dependent~\cite{Accardi:2009qv}.
We find this makes a very strong case for future nuclear TMD 
experiments, providing a completely independent measurement 
of such an important nuclear property.

Finally, we have seen that even though not many data are available
at present, an important experimental effort is ongoing at JLab, both to
analyze existing data and to perform new experiments. We can expect important
experimental progresses with the 12~GeV upgrade of JLab, on both nuclear
GPDs and TMDs of light nuclei. Further in 
the future, the construction of an electron ion collider~\cite{Accardi:2012qut}
would allow to perform many of the measurements discussed here, with high 
precision and wide kinematic coverage.

\bibliography{mybib}{}
\bibliographystyle{epj}

\end{document}